\documentstyle[twocolumn]{article}
\def\v{\vskip 3mm}
\def\h{$^{\rm h}$}
\def\m{$^{\rm m}$}
\def\s{$^{\rm s}$}
\def\HI{H {\sc i}}
\def\HII{H {\sc ii}}
\def\c{\centerline}
\def\kms{ km s$^{-1}$ }
\def\co{$^{12}$CO ($J=1$--0)}
\def\vlsr{$V_{\rm lsr}$}
\def\Msun{M_{\odot \hskip-5.2pt \bullet}}

\begin{document}

\title{The Virgo High-Resolution CO Survey. III. -- NGC 4254 --}

\author{Yoshiaki Sofue,$^1$  Jin Koda,$^{1, 2}$ Hiroyuki Nakanishi,$^1$
 and Makoto Hidaka$^1$ \\
{\it $^1$ Institute of Astronomy, University of Tokyo, Mitaka, Tokyo 181-0015}\\
{\it $^2$ Nobeyama Radio Observatory, National Astronomical Observatory, 
Mitaka, Tokyo 181-8588}\\
{\it E-mail sofue@ioa.s.u-tokyo.ac.jp}}
\date{(Received 2002 August 15; accepted 2002 October 15)}
\maketitle

\begin{abstract}
We present high-angular-resolution ($1''.5$--$5''$)  interferometer
observations of the \co\ emission in the central region of the SA(s)c
galaxy NGC 4254.
The observations were obtained using the Nobeyama Millimeter-wave Array
(NMA) during the course of a long-term CO line survey of Virgo spirals.
We present the spectra, channel maps, integrated intensity distributions,
velocity fields, position--velocity diagrams, and compare the data with
various optical images.
The rotation velocity is already finite at the nucleus, or at least it
rises steeply to 80 \kms\ within the central 1$''$, indicating the
existence of a massive core of $10^8 \Msun$ within $1''$ (80 pc) radius.
The CO intensity maps show that the inner disk has well-developed multiple
spiral arms, winding out from a bar-shaped elongated molecular complex.
In addition to the bisymmetric spiral arms, an asymmetric tightly wound arm
with high molecular gas density is found to wind out from the molecular bar.
The molecular spiral arms, particularly the tightly wound arm, well traces
optical dark lanes, and are associated with H$\alpha$ arms having many 
H~{\sc ii} regions.
The inner asymmetric spiral structures can be explained by ram-pressure
distortion of inter-arm low density regions of the inner disk by the
intra-cluster gas wind, and is indeed reproduced by a hydrodynamical
simulation.
\end{abstract}

\section{Introduction}

The Virgo member spiral NGC 4254 (M99) is an SA(s)c galaxy (RC3) with an
optical radius of $5'.36$ (de Vaucouleurs et al. 1991), and an inclination angle
of $42^{\circ}$ (Phookun et al. 1993).
We adopted the Cepheid calibrated distance of 16.1 Mpc of the
Virgo Cluster for this galaxy (Ferrarese et al. 1996), with which
the Tully--Fisher distance of 16.8 Mpc is consistent
(Sch\"oniger, Sofue 1997).
The parameters of the galaxy are given in table 1.
Optical images show that the galaxy is dominated by a three-arm structure,
exhibiting $m=1$ and $m=3$ modes (Iye et al. 1982; Gonz\'alez, Graham 1996).
This one-armed structure is also seen in the \HI\ gas distribution
(Phookun et al. 1993).
Such an asymmetric spiral pattern is often observed in tidally interacting
galaxies, but there is apparently no massive companion around NGC 4254.
Moreover, a tidal interaction generally produces a bisymmetric structure with
near and far side arms, and the resultant spiral structure shows $m=2$ mode
features, but not the $m=1$ or $m=3$ mode.
Therefore, the asymmetric spiral pattern of NGC 4254 would be due to some
other mechanisms, such as an interaction with the intracluster medium (ICM),
as suggested by the H {\sc i} head--tail structure.
It is also an interesting question as to 
how deeply the ICM ram effect can affect
the galactic disks and whether the molecular disks are indeed disturbed.
In fact, Hidaka and Sofue (2002) have shown by a numerical simulation that 
a ram effect by ICM wind can produce lopsided molecular arms in the central 
few kpc region, if the inter-arm gas density is low enough even though the
mean radial density is high enough to avoid the ram effect.
In such a circumstance, the ICM ram pressure perturbs the orbits of the
low-density inter-arm gas, which results in significantly displaced shocked
arms from unperturbed arms.

In this paper, we present high-resolution \co\ line observations of
NGC 4254 using the Nobeyama mm-wave Array (NMA) in AB, C and D array
configurations.
The observational parameters and detailed procedures are described in
the first paper of this series, reporting the high-resolution
CO line Virgo survey (VCOS: Sofue et al. 2003a).
We discuss the kinematics and ISM properties based on the CO data,
and compare the result with various optical images.
Environmental effects, such as the tidal interaction and/or the ram-pressure
effects, would be an interesting subject for some Virgo galaxies located 
near to the cluster center and moving in the dense intracluster medium.
Based on the CO data, we discuss the possibility that the distribution and
kinematics of molecular clouds in the central region can be affected by the ram-pressure effect of the intracluster medium.

\begin{table*}
\begin{center}
\caption{Properties of observed galaxies}
\vskip 2mm
\begin{tabular}{cc}
\hline\hline
Morphology$^1$ & SA(s)c  \\
NED position$^2$ (J2000)&$\alpha$ = 
	12\h18\m49.56\s\\ 
	& $\delta$ = $+14^\circ24'59''.4$ \\
Apparent magnitude$^1$ & 10.44  \\
Systemic velocity$^1$  & 2407 km sec${}^{-1}$  \\
Inclination angle$^3$  & 42$^\circ$     \\
Position angle$^3$ & $68^{\circ}$  \\
Assumed distance$^4$  & 16.1 Mpc ($1'' = 78.06$ pc)   \\
\hline
\end{tabular}
\end{center}
References: 1. de Vaucouleur et al. (1991);
2. NASA Extragalactic Database (http://nedwww.ipac.caltech.edu/);
3. Phookun et al. (1993);
4. Ferrarese et al. (1999).
\end{table*}

\section{Observations}

\co high-angular-resolution interferometer observations of the central
region of NGC 4254 were carried out on 2000 March 10--11, 2000 February
4--5, and 1999  December 8, using the Nobeyama Millimeter Array (NMA) 
in the AB, C, and D configuration, respectively.
Tables 2 gives the observational parameters.
Each observation run took typically 8 hours, including calibrations.
The pointing center was at
RA (J2000)=12\h18\m50.03\s,
Dec (J2000)=$+14^\circ24'52''.8$,
adopted from Condon et al. (1990).
However, this position was found to be $5''.9$ to the east and $6''.3$ to the
south of the newly determined kinematical center of the galaxy using our
high-resolution CO-line velocity field.
In this work, therefore, we adopted the new kinematical center as our origin
of the maps, whose coordinates are
RA (J2000) = 12\h18\m49.61\s,
Dec (J2000)=  $+14^\circ24'59''.1$.
The present center position may be compared with the other observations
listed in table 3, which all coincide within $\sim \pm 0''.5$.

\begin{table*}
\begin{center}
\caption{Observation Parameters.}\vskip 2mm
\begin{tabular}{lc}
\hline\hline
Observed center frequency  & 114.353 GHz \\
Array configurations & AB, C, and D  \\
Observing field center (Condon et al. 1990): &    \\
 \hskip 10mm $\alpha$ (J2000) & 12\h18\m50.03\s\\
 \hskip 10mm $\delta$ (J2000) & $+14^\circ24'52''.8$\\
Derived dynamical center: \\
\multicolumn{2}{c}(assumed to be nucleus position $\simeq$ map center in this paper)\\
 \hskip 10mm $\alpha $ (J2000) & 12\h18\m49.61\s \\
 \hskip 10mm $\delta $ (J2000) & $+14^\circ24'59''.1$ \\
Central velocity  & 2400 \kms  \\
Frequency channels & 256  \\
Total bandwidth  & 512 MHz  \\
Velocity coverage  & 1342 \kms  \\
Velocity resolution  & 5.24 \kms  \\
Amplitude and phase calibrator & 3C 273  \\
Primary beam & 65$''$  \\
Cell size & 0$''$.25 \\
\hline
\end{tabular}
\end{center}
\end{table*}

\begin{table*}
\begin{center}
\caption{Comparison of our center position with other observations.}
\begin{tabular}{lccc}
\hline\hline
RA (J2000) & Dec (J2000) & References \\
\hline
12\h18\m49.61\s &  $+14^\circ24'59''.1$ &CO: This work \\
12\h18\m49.63\s &  $+14^\circ24'58''.8$ &CO: Sakamoto et al. (1999) \\
12\h18\m49.43\s & $+14^\circ24'59''.5$ & $B$-band photometry: 
Yasuda et al. (1995)\\
12\h18\m49.73\s &  $+14^\circ24'58''.8$ &IRAS: Soifer et al. (1987)\\
\hline
\end{tabular}
\end{center}
\end{table*}

\begin{table*}
\begin{center}
\caption{Parameters of Maps$^\dagger$.}
\begin{tabular}{cccccccccc}
\hline\hline
  & &  \multicolumn{2}{c}{Beam} & \multicolumn{2}{c}{Velocity} &
  & r.m.s. noise
  &$T_{\rm b}$ for \\
  \cline{3-4} \cline{6-7}
Resolution & Weighting &  FWHM  & P.A.&&  Resolution & Sampling
&$\sigma$ & 1 $\rm Jy\,beam^{-1}$\\
  &  &(arcsec)  & (deg)  &&($\rm km\,s^{-1}$)
  &($\rm km\,s^{-1}$) & ($\rm mJy\,beam^{-1}$) & (K) \\
\hline
Low  & Taper & $5.22 \times 3.45$   & 152.0  &&  21.0 & 10.5 &  19.6 & 5.1 \\
Medium & Natural & $2.99 \times 2.34$ & 148.0  &&  21.0 & 10.5 &  15.4 & 13.2\\
High  & Uniform  & $1.67\times1.53$   & 167.5  &&  21.0 & 10.5 &  21.3 & 36.0\\
\hline
\end{tabular}\\
\end{center}
$\dagger$ The map centers are set at the derived dynamical center:\\
($\alpha_{2000}$, $\delta_{2000}$) =(12\h18\m49.609\s, $+14^\circ24'53''.1$)
\end{table*}

We also observed nearby radio point source 3C 273 as a flux and phase
calibrator every 20 min.
Because the intrinsic flux density of 3C 273 at the observing frequency was
non-periodically variable, we performed flux calibration for each
observation.
The flux density of 3C 273 was 11.25, 12.40, and 11.26 Jy, in the observations
of the AB, C, and D configurations, respectively.
We used a spectro-correlator system, Ultra Wide Band Correlator (UWBC: Okumura
et al. 2000) in a narrow-band mode, which had 256 channels;
the total bandwidth was 512 MHz.
One channel corresponded to 5.24 \kms\ at the observing frequency.

The raw data were calibrated using UVPROC-II, a first-stage reduction system
developed at the Nobeyama Radio Observatory (NRO), and were then
Fourier-transformed using the NRAO Astronomical Image Processing System (AIPS).
We reduced the thus-obtained dirty map by the CLEAN method using three 
different weighting functions and tapering, as summarized in table 4.

First, we obtained low-resolution maps using a natural weighting function,
tapered by a Gaussian function having deviations of 60 k$\lambda$ and a
cut-off at greater than 80 k$\lambda$ in $(u,v)$ space.
The data were averaged in 4 bins, yielding a 128-channel data cube with
a velocity resolution of 21 \kms; the channel increment was 2, corresponding
to 10.4 \kms.
The synthesized beam was $5''.22\times 3''.45$, and the obtained channel had
a typical r.m.s. noise of 19.6 mJy/beam.

Second, we obtained the most representative maps from the present
observations using a natural weighting function, no taper.
The data were averaged in 4 bins (21 \kms) of the original channels, and the
channel increment was 2 (10.4 \kms), yielding a 128-channel data cube.
The synthesized beam was $2''.99\times 2''.33$, and the typical r.m.s. noise
on a channel map was 15.4 mJy/beam.

Third, we obtained high-resolution maps, using a uniform weighting function,
no taper.
The data were averaged in 2 bins giving a velocity resolution of 10.4 \kms;
the channel increment was 2 (10.4 \kms).
The synthesized beam was $1''.67\times 1''.53$, and the typical r.m.s. noise
was 21.3 mJy/beam.

\section{Results}

\subsection{Spectra}

Figures 1a and b show CO line spectra averaged in a $1'\times1'$ squared
region (4.89 kpc square) around the center and a $20''\times20''$ (1.69 kpc)
squared region, respectively.
Both spectra indicate double horn shapes, typical for a rotating disk.
However, the central spectrum in figure 1b has a much narrower width,
$\sim 130$ \kms, than that of the outer region, where the width is 210 \kms.
Since the double-horn feature is typical for a rotating disk with a constant
rotation velocity, the clearly different velocity widths suggest that the disk
rotation velocity varies step-like at a few kpc from the center, which is
discussed  in a more detail using position--velocity diagrams in a later
section.

Figure 1c shows a CO line spectrum of NGC 4254 obtained by convolution with a
synthesized Gaussian beam of FWHM $45''$ in order to compare the present CO
intensity with that of the FCRAO 14-m single-dish observation
(Kenney, Young 1988).
The peak intensity and the integrated intensity of our NMA observation are
2.39 Jy/Beam and 325.58 Jy/Beam \kms, corresponding to 106 mK and 14.8 K \kms,
respectively.
A FCRAO CO observation of NGC 4254 shows that the peak intensity is
$ 89 /\eta_B$ mK, and the integrated intensity is 
$(10.6 \pm 2.0)/\eta_{\rm B}$ K \kms, 
where $\eta_{\rm B}$ is the beam efficiency of the 
FCRAO 14m telescope, which is equal to $0.53 \pm 0.04$.
Hence, the integrated intensity of our NMA observation covers $74 \%$ of the
integrated intensity of the FCRAO single-dish observation.

\v
\c{--- Fig. 1a, b, c ---}
\v

\subsection{CO Intensity Maps}

Figure 2 shows channel maps at a velocity separation of 10.4 \kms.
Each channel map displays an averaged brightness in a 21 \kms\ velocity range.
The CO emission is visible from the 6th channel at \vlsr\ = 2300 \kms to the
27th channel at 2517 \kms.
The distributions in individual channel maps are generally along iso-velocity
lines, typical for a rotating disk galaxy, while they are patchy, indicating
that the distribution is not uniform, but concentrated in the arms.

\v
\c{--- Fig. 2 ---}
\v

Figure 3 shows the total CO intensity map at a low resolution of $5''.22\times
3''.45$.
Figure  4a shows the total CO intensity map at the representative resolution
of $2''.33 \times 2''.99$ using a natural weighting function.
No primary beam correction has been applied.
The major and minor axes across the dynamical center are shown by the
big crosses in figures 3 and 4a.
Figure 5a shows the same view but at the highest resolution  of
$1''.67\times 1''.53$ with uniform weighting.

\v
\c{--- Fig. 3 ---}

\c{--- Fig. 4a, b, c---}

\c{--- Fig. 5a, b, c ---}
\v

\subsubsection{Central CO bar}
The CO emission shows an elongated bar-like concentration around the map
center (kinematical center) with the major axis at a P.A. of about 50$^\circ$,
being displaced from the galaxy's major axis at P.A. = 68$^\circ$.
The center of gravity of this bar is slightly offset from the kinematical
center toward the northwest.
Although the CO distribution shows a bar feature, the optical morphology
is non-barred SA(s)c type (RC3).
In fact, no bar feature is recognized in either the R-band image shown
in figure 6a or in K-band image in figure 8, as is shown later.

\subsubsection{Spiral arms}
Two major spiral arms are winding out from both ends of this central
molecular bar in an counterclockwise direction, and extend until the field edge.
In figure 3 we name the western major arm Arm I, and  the eastern major
arm Arm II.
At RA=12\h18\m51.5\s, Dec=$14^\circ24'10''$ (J2000), there is a segment of an 
arm-like feature with a dense molecular complex, running parallel to Arm II.

\subsubsection{Tightly wound arm}
Soon after it starts from the eastern end of the central CO bar, Arm II
bifurcates into a more tightly wound dense arm, which we call Arm III,
as indicated in figure 3.
The western end of this tightly-wound arm runs closely parallel to Arm I, as
is more clearly seen in figures 4a and 5a.
This tight arm has a much higher brightness, about twice to three times
that of the two major spiral arms.

\subsubsection{North--South Asymmetry}
The global distribution of molecular gas in the observed region of NGC 4254,
as shown in figures  3 to 5, is highly asymmetric with respect to the major
axis. The southern half is much CO brighter compared to the northern half.
Most of the asymmetry comes from the tight CO arm, Arm III.

\subsection{Velocity Field}

Figure 4b shows an intensity-weighted velocity field corresponding to
figure 4a with natural weighting, and figure  4c is an overlay of the same
velocity field on the integrated intensity map at a lower resolution in the gray
scale, which is the same as figure 3. Figure 5c shows the  velocity field for
the central region at high resolution, corresponding to figure  5b.
Figure 5d enlarges the central $4''$ region of figure  5c.

We determined the kinematical center as the position where the 
iso-contours run most tightly in the high-resolution velocity field in 
figures  5c and d:
RA (J2000) = 12\h18\m49.61\s\ and
Dec (J2000) = $+14^\circ24'53''.1$
with an accuracy of $\pm 0''.3$.
We adopt these coordinates as the center position of the galaxy, and assume
that the position coincides with the nucleus.
These coordinates are in accordance with the dynamical center determined by
applying the task GAL in the AIPS reduction package, which uses a Brandt
rotation curve to fit the (RA, Dec, Velocity) cube. The fitted result to
the entire disk in the observed area was
RA (J2000) = 12\h18\m49.56\s, Dec (J=2000) =$+14^\circ24'58''.5$.
The GAL fitting also gave the position angle of the molecular disk
to be P.A. = $66^\circ.5$, inclination $38^\circ.8$, and the systemic velocity
2403.6 \kms, which are consistent with the adopted parameters in table 1.

The general pattern of the velocity field in figure  4b shows a symmetric
spider diagram, indicating a regular circular rotation of the CO disk,
with the eastern half red-shifted and the western half blue-shifted.
Assuming that the spiral arms are trailing, the rotation is clockwise and,
accordingly, the northern side is the near side.
On smaller scales, the non-circular velocity components are superimposed.
Individual spiral arms show iso-velocity contours running obliquely to those
expected from a circular rotation, indicating non-circular streaming motion
due to spiral density waves.

The central CO bar, whose major axis is at about P.A. = $50^\circ$, is rotating
rather circularly with the node of the velocity field coinciding with that of
the P.A.  of the galaxy at $68^\circ$.
However, the very central region within $2''$ from the nucleus is superposed
by non-circular motion, where the iso-velocity contours run in an
integral-sign shape with their average direction in the north-south, and the
velocity node is at  P.A. = $80^\circ$, significantly displaced from the
galaxy's major axis.

\subsection{Overlays on Optical Images}

In figure  6a we overlay our CO intensity contour map at medium resolution
(figure  4a) on a $g$-band image from the Princeton Galaxies Catalogue obtained
with the Palomar 1.5-m telescope by Frei et al (1996).
The kinematical center was assumed to coincide with the optical nucleus.
Scaling and positioning are accurate within an error of about $2''$, which
applies to all the following overlays.
Figure  6b shows the same region, but superposed on a 5957A-band image 
from the HST WFPC2 archive.
Figure  6c shows the central region with the high-resolution CO contour map
on the HST image as in figure  6b.
The CO arms, particularly the tightly-wound dense CO arm (Arm III),
well trace the optical dust lanes running along the inner sides of the
optical spiral arms.
An inner dust lane at about $6''$ south-east of the nucleus is also traced by
a CO arm, starting from the eastern mid-point of the central CO bar.
Inside the CO bar, most of the CO clumps are generally associated with dust
lanes and/or dusty clumps.
However, the molecular bar, itself, does not well draw an overall spiral
pattern, whereas the HST optical image within $10''$ of the nucleus shows
symmetric amorphous spiral patterns.

Figure 6d shows an overlay of a medium-resolution CO map on a negative
H$\alpha$ image taken from the NED archive of a survey of \HII\ regions in
spiral galaxies obtained by Banfi et al (1993) using the 2.1-m KPNO telescope.
This figure demonstrates an excellent global correlation of molecular arms and
clouds with \HII\ regions, indicating that star formation occurs in spiral arms
with dense molecular clouds.
A more detailed inspection of figure  6d reveals, however, that some intense
molecular clouds are not directly superposed on \HII\ regions, but are slightly
displaced from the star-forming region.
Also, some strongest \HII\ regions are not superposed by CO clumps, but are
slightly shifted from molecular cloud centers.
The displacement generally occurs in the direction perpendicular to the arms
in the sense that that the \HII\ regions are located outside of the CO arms
(dark lanes). This is reasonably explained by the galactic shock-wave theory.
However, the displacement also occurs among \HII\ regions and molecular clumps
along the arms.
This fact shows that intense molecular clouds are places where star formation
has not yet started.
On the other hand, the strongest \HII\ regions have already exhausted 
their parent molecular clouds.

\v
\c{--- Fig. 6a, b, c, d ---}
\v

\subsection{PV diagram}

The position--velocity (PV) diagrams along the major axis at P.A.$=68^\circ$
crossing the nucleus of NGC 4254 are shown in figures  7a to c in the order
of low, medium, and high resolutions, respectively.
The slit widths are $10''$, $3''$ and $2''$, respectively.
Figure 7c enlarges the central part.

Emission in the PV diagram is separated into two major concentrations around
the nucleus at finite velocity offsets of about $\pm 40$ \kms\ from the
systemic velocity.
The PV diagram shows a clear intensity minimum at the nucleus,
which is prominent in figure  7c,  despite the continuous distribution of
CO emission in the integrated intensity maps (figures  3 to 5).
This indicates that the rotation velocity rises very steeply within the
central $1''$ or less.

After attaining the intensity maximum in the PV diagram, the velocity
increases more gently toward the edges of the bar.
Near the ends of the bar, the PV velocity increases step-like, and attains
maximum velocities of about $\pm 100$ \kms\ in the spiral arms.

\v
\c{--- Fig. 7a, b, c ---}
\v

\section{Discussion}

\subsection{Rotation Curve and Dynamical Mass}

The inclination angle of NGC 4254 was obtained to be $i=42^{\circ}$ from 
\HI\ data (Phookun et al. 1993).
Figure 8 shows a contour-form $K$-band image taken from NED data archive of the
near-IR surface photometry of spiral galaxies (M\"ollenhof,  Heidt 2001).
We determined the inclination angle by an ellipse fit to the iso-intensity
contours of the $K$-band image, and obtained an inclination angle of
$i=42 (\pm1)^\circ$, consistent with the earlier value.
We adopted this inclination angle in the present work.
On the other hand, more face-on values have been obtained from CO
observations: $29^{\circ}$  (Sakamoto et al. 1999) or $28^{\circ}$
(Kenney, Young 1988).
Our CO maps, representing the two open spiral arms, also suggest a more
face-on value.
However, we may rely more on the inclinations from NIR images, because the CO
maps manifest spiral-shocked gaseous arms, but not necessarily the backbone
stellar disk.

\v
\c{--- Fig. 8 ---} 
\v

We used PV diagrams to obtain a rotation curve by applying an iteration
method developed by Takamiya and Sofue (2002) and Sofue et al (2003b).
Figure 9a shows the obtained rotation curve with an inclination angle of
$42^\circ$ being corrected.
The rotation velocity rises steeply in the central $\sim 100$ pc,
and reaches to 120 \kms\ maximum, followed by a small dip at 250 pc.
It then rises to a maximum velocity of 190 \kms at $r\sim 900$ pc, and
declines to 150 \kms at 1.5 kpc.
This maximum may be due to the central bulge.
Then, the rotation velocity gradually increases, representing the
disk component.
The small-scale fluctuation of amplitudes by $\sim \pm 10$ \kms\ may not be
real and, hence, the small local dips should not be taken seriously.

The dynamical mass in the central 100 pc is $3\times 10^8\Msun$ for a rotation
velocity of 120 \kms,  and the mass within 1 kpc is $7\times10^9 \Msun$ for
180 \kms.
The high central velocity and massive core within the 100 pc region of the
nucleus was observed for many galaxies (Takamiya, Sofue 2000;
Sofue et al. 2001; Koda et al. 2002).

\v
\c{--- Fig. 9a, b ---} 
\v

We calculated the surface-mass distributions (SMD) using the observed
rotation curve by applying a deconvolution method described in
Takamiya and Sofue (2000).
Figure 9b shows the derived mass distributions, where the dashed line was
calculated by a spherical-symmetry assumption, and the full line by a flat-disk
assumption.
Since the rotation curve is given only within 3 kpc, the results beyond
2.5 kpc are not real.
Both the dashed and full-line results agree with each other, except for
the larger fluctuations and slightly larger values for the spherical assumption.
The figure indicates a high-density central peak, representing the massive core
of scale radius of 80 pc with the central value being as high as
SMD = (2--3) $\times 10^4 \Msun{\rm pc}^{-2}$.
Since the values are resolution-limited, the real core radius would be smaller
and the density would be higher.
The massive core is followed by an exponentially decreasing part at
$r\sim$ 0.2--1 kpc with a scale radius of 600 pc, likely representing the
bulge component.
Beyond 1 kpc, the SMD decreases more slowly, representing an exponential
disk component of the scale radius of about 3 kpc, though the accuracy is
much poorer for this component, because of the small radius of deconvolution.

\subsection{Molecular Gas Mass}

The total CO luminosity mass within the inner $22''.5$ radius region is
estimated to be 2.4 $\times 10^8 \Msun$, using a conversion factor of
$C = 2.1 \times 10^{20}$ cm${}^{-2}$ K${}^{-1}$ km$^{-1}$ s (Arimoto et
al. 1996).
Taking into account the missing flux of our NMA observation and that the mass
of molecular contents including He and other elements is 
$M_{\rm{gas}} = 1.36 M_{\rm{H}_2} $, the total mass of molecular gas is
calculated to be $4.34 \times 10^8 \Msun$.
The CO luminosity mass within the inner $5''$ radius region, where a missing
flux correction is not needed, is $4.0\times 10^7 \Msun$.
The luminosity mass to dynamical mass ratio in the inner $5''$ is 0.15, much
larger than that in the inner $22''.5$, where the ratio is 0.026.
If we adopt a more face-on value for the inclination, e.g. 29$^\circ$, the
above ratios should be decreased by a factor of 0.52.
Hence, the molecular-gas mass in NGC 4254 is not particularly large compared
with the dynamical mass.

\subsection{Origin of the North--South Asymmetry in CO}

Environmental effects, such as the tidal interaction and/or the ram-pressure
effects, are an interesting subject concerning the Virgo galaxies.
NGC 4254, which is apparently not associated with a companion galaxy, is known
for its distorted \HI\ structure, and the lopsidedness is supposed to be the
result of an environmental effect due to the ram pressure by the intracluster
medium (Phookun et al. 1993).
The outer \HI\ gas is largely extended in the northeastern area over
$2''~ (\sim 10$ kpc), and is associated with several bifurcated optical arms.
One prominent optical/\HI\ arm extends from the south-to-western region;
this prominent one-armed feature leads to the $m=1$ mode based on a spiral mode
analysis (Iye et al. 1982).

It was long believed that the inner disks are stable and not disturbed by
such an environmental effect as the ram pressure (Vollmer et al. 2001).
However, simulations have already shown that the ram effect is significant,
even on the inner disks (Sofue 1994).
Recently, we examined the ram effect on the inner molecular disk in
detail, and showed that it is significant when it acts on the
inter-arm regions, where the gas density is much lower than in the arm
regions, and hence the ram-pressure easily disturbs the orbits of the inter-arm
gas (Hidaka, Sofue 2002). The disturbed orbits of inter-arm gas leads to
a significant displacement of the shocked arms from the regular bisymmetric
arm positions, resulting in distorted inner molecular spiral structures.
Figures 10a (top right) and 10b (top left) show the result of our
two-dimensional hydrodynamical simulation of ram pressure on a gas disk with
spiral arms.
Here, the intracluster wind blows from the west toward the east at an inflowing
angle of 45$^\circ$ to the galactic plane with a wind velocity of 1000 and
1500 \kms, respectively, and the ICM density is assumed to be
$5\times 10^{-4}$ H cm$^{-3}$.
A detailed description of the model and simulation procedure is given in Hidaka
and Sofue (2002).

\v
\c{--- Fig. 10a, b, c ---}
\v

The simulation may be compared with the observed CO intensity distribution,
shown in figure 10c (bottom), the same as figure 4a.
The distribution of molecular gas in the central $1'$ region of NGC 4254 is
extended to the south-eastern half, where the tightly wound CO arm is most
prominent.
This tight arm winds out to the north, followed by the NE optical outer arms.
The western arm winds out to the south, and continues to the optical/\HI\
one-armed spiral arm.
These asymmetric arm structures, particularly the tight CO arm (Arm III),
are well mimicked by the simulation in figures 10a and b, although the
details are not necessarily reproduced.

\vskip 5mm

We are grateful to the staff at NRO for their helpful discussion about
the observation and reduction.
The optical images were taken via the NASA Extragalactic Data
Archive (NED), and from the Hubble-Space-Telescope Data Archive at STScI
operated by NASA.
J.K. was financially supported by the JSPS (Japan Society for the Promotion
of Science) for young scientists.

\vskip 10mm

\parskip=0pt
\def\r{\hangindent=1pc \noindent}
\def\bibitem{\r}
\noindent{\bf References}

\bibitem Arimoto, N., Sofue, Y., \& Tsujimoto, T. 1996, PASJ, 48, 275

\bibitem Banfi, M.,  Rampazzo, R., Chincarini, G., \& Henry, R. B. C. 
1993 A\&A 280, 373 

\bibitem Condon, J. J., Helou, G., Sanders, D. B., \&
Soifer, B. T., 1990, ApJS, 73, 359

\bibitem de Vaucouleurs, G., de Vaucouleurs, A., Corwin, H. G. Jr., 
Buta, R. J., Paturel, G., \& Fouqu\'e P. 1991, 
Third Reference Catalog of Bright Galaxies (New York: Springer-Verlag)

\bibitem Ferrarese, L., Freedman, W. L., Hill, R. J., Saha, A.,  Madore, B. F.,
 Kennicut, R. C. Jr., Stetson, P. B., ford, H. C., et al. 1996 ApJ, 464, 568

\bibitem Frei, Z., Guhathakurta, P., Gunn, J. E.,  \& Tyson, J. A., 1996,
AJ, 111, 174

\bibitem Gonz\'alez, R. A.,  \& Graham, J. R. 1996, ApJ, 460, 651

\bibitem Hidaka, M., \& Sofue Y. 2002, PASJ, 54, 33

\bibitem Iye, M., Okamura, S., Hamabe, M., \& Watanabe, M. 1982, ApJ, 256, 103

\bibitem Kenney, J. D., \&  Young, J. S., 1988, ApJS, 66, 261

\bibitem Koda, K.,  Sofue, Y., Kohno, K., Nakanishi, H.,  Onodera, S.,
Okumura, S. K.,  \& Irwin, Judith A. 2002, ApJ, 573, 105

\bibitem M\"ollenhoff, C.,  \& Heidt, J. 2001, A\&A, 368, 16

\bibitem Okumura, S. K.,  Momose, M., Kawaguchi, N., Kanzawa, T.,
Tsutsumi, T., Tanaka, A., Ichikawa, T., Suzuki, T., et al. 2000, PASJ, 52, 339

\bibitem Phookun, B., Vogel, S. T., \& Mundy, L. 1993, ApJ, 418, 113

\bibitem Sakamoto, K., Okumura, S. K., Ishizuki, S., \& Scoville, N. Z.,
1999, ApJS 124, 403

\bibitem Sch\"oniger, F., \& Sofue, Y., 1997, A\&A, 323, 14

\bibitem Sofue, Y. 1994, ApJ, 423, 207

\bibitem Sofue, Y., Koda, J., Kohno, K., Okumura, S. K., Honma, M.,
Kawamura, A., \& Irwin, Judith A.  2001 ApJ, 547,  L115

\bibitem Sofue, Y., Koda, J., Nakanishi, H., \& Onodera, S.
 2003b PASJ, in this volume.

\bibitem Sofue, Y., Koda, J., Nakanishi, H., Onodera, S. Kohno, K., 
Tomita, A., \& Okumura, S. K.  2003a PASJ, in this volume.

\bibitem Soifer, B. T., Sanders, D. B., Madore, B. F., Neugebauer, G.,
Danielson, G. E., Elias, J. H., Lonsdale, C. J., \& Rice, W. L. 1987, 
ApJ, 320, 238

\bibitem Takamiya, T, \& Sofue, Y., 2000, ApJ 534, 670

\bibitem Takamiya, T, \& Sofue, Y., 2002, ApJ, 576, L15

\bibitem  Vollmer, B., Cayatte, V., Balkowski, C., \& Duschl, W. J.
2001, ApJ, 561, 708

\bibitem Yasuda, N., Okamura, S., \& Fukugita, M. 1995, ApJS, 96, 359

\newpage

\parindent=0pt
\parskip=4mm
\noindent Figure Captions
\vskip 5mm
PS figures are available from http://www.ioa.s.u-tokyo.ac.jp/radio/virgo/
\vskip 5mm

Fig. 1 (a) Spectrum averaged in the central $60''$ squared area, showing a
double-horn shape indicative of a rotating disk.

(b) Spectrum averaged in the central $20''$ squared area, showing a
double-horn shape, but with a much narrower width than in figure  1a, indicative
of a rotating disk at a slower velocity.

(c) Spectrum at the center after convolved with a Gaussian beam of FWHM
45$''$.

Fig. 2. Channel maps of the CO-line intensity at a 10.4 \kms\ 
velocity increment,
each indicating the CO-line intensity integrated in a 21 \kms\ velocity range.
The 10th, 15th and 20th channels correspond to $V_{\rm lsr}$ = 2340.2, 2392.6,
and 2444.2 \kms, respectively.
The intensity unit is the brightness temperature in K.
Contours are drawn at every 20\% of the peak intensity of 3.93 K.

Fig. 3. Integrated CO-line intensity map (the zero-th moment map) of NGC 4254.
The $uv$ weighting function is natural, And tapered.
The synthesized beam is $5''.22 \times 3''.45$ (P.A. $ = 152^{\circ}$).
The contour levels are every 10\% of the peak intensity of
85.2 K \kms (16.6 Jy beam${}^{-1}$ km sec${}^{-1}$).
The major and minor axes are indicated by the cross.

Fig. 4
(a) Integrated CO intensity map. The $uv$ weighting function is natural,
and non--tapered.
The synthesized beam is $2''.99 \times 2''.34$  (P.A. $ = 148^{\circ}$).
The central cross indicates the dynamical center and the major and minor axes.
Because the attenuation due to the primary beam pattern
is not corrected,  the noise level is uniform in this image.
The contour levels are every 10\% of the peak flux of
110.3 K \kms (8.33 Jy beam$^{-1}$ \kms).
The r.m.s. noise in the map is $1\sigma = 388$ mJy beam$^{-1}$ \kms.

(b)  Velocity field of NGC 4254 (first moment map) corresponding to figure 4a.
The synthesized beam is equal to that of figure  4a.
The interval of the iso-velocity contours is 10 \kms.

(c) Overlay of the velocity field on the low-resolution intensity distribution
 from figure 3 in gray.

Fig. 5 (a) Integrated CO intensity map of NGC 4254, with uniform weighting.
The synthesized beam is $1''.68 \times 1''.53$ (P.A. $ = 167^{\circ}.5$).
The contour levels are every 10\%  of the peak flux of 205.3 K \kms
(5.694 Jy beam$^{-1}$ \kms).

(b) Enlargement  of the central part of figure 5a. The beam size and contour
intervals are the same as figure 5a.

(c)  Velocity field of NGC 4254 (first moment map) corresponding to figure
5a.  The synthesized beam is equal to that of figure  5a. The interval of
the iso-velocity contours is 10 \kms.

Fig. 6 (a) Overlay of the integrated CO intensity map, same as figure 5a, on
an $R$-band negative image of NGC 4254 reproduced from Frei et al.'s (1996)
galaxy catalog.  
(b)  Overlay of the integrated CO intensity map, same as figure 5a, on the
B-band positive image of NGC 4254, as reproduced from the galaxy-image
archive of the Hubble-Space-Telescope.  
(c)  The same as figure  6a, but enlarged for the central part.  
(d)  The same as figure  6a, but overlaid on an H$\alpha$ image from Banfi
et al. (1993) taken from the NED archive.

Fig. 7 (a) Position-velocity diagram of the CO emission in NGC 4254 from the
tapered cube ($5''.22 \times 3''.45$ beam, velocity resolution 21 \kms) along
a $10''$-width slit on the major axis (P.A. $= 68^{\circ}$) across the
kinematical center.
the contour levels are at 5, 10, 20, .... 100\% of the peak value
of 1.73 K (0.349 Jy beam${}^{-1}$).
The velocity centroid is 2402.6 \kms.
(b) Same as figure  7a, but for natural weighting and a non--tapered cube
($2''.99 \times 2''.34$  beam, and velocity resolution 21 \kms) along a
$3''$-width slit on the major axis.
The contour levels are at 5, 10, 20, 30, ... 100\% of the peak value of
3.23 K (0.248 Jy beam${}^{-1}$).
(c)  Same as figure  7a, but for the central part from the
high-resolution cube ($1''.68 \times 1''.53$ beam, and velocity
resolution of 10.4 \kms) in a $2''$-width slit along the major axis.
The contour levels are every 10\% of the peak value of 5.46 K
(0.152 Jy beam${}^{-1}$).

Fig. 8. $K$-band image of NGC 4254 in a contour form, taken from the NED archive
of the NIR survey of nearby galaxies by M\"olenhoff and Heidt (2001).
The displayed area is $4'\times 4'$ around the nucleus, and 1 pixel corresponds
to $1''$.
The contours are drawn at 2, 4, ... 12, 15, 20, ...  30, 40, ... 100\% of
the peak value of 7.34 DN/pixel.

Fig. 9 (a) Rotation curve for the inner 3 kpc of NGC 4254 obtained by the
iteration method using the PV diagram, as described in Takamiya and Sofue
(2002) and Sofue et al. (2003b).

(b) Surface mass distribution (SMD) obtained by deconvolution of the rotation
curve using the method described in Takamiya and Sofue (2000).
The full line represents the result for a flat-disk assumption, and the
dotted line for a spherical-symmetry assumption of the mass distribution.
Note the three components: the central massive core at 0--100 pc
with a scale radius of 80 pc, a bulge at 0.1--1 kpc, and the disk component at
1 to 2.5 kpc.

Fig. 10. Two-dimensional hydrodynamical simulation of the effect of ram
pressure by an intracluster wind on the inner part of a spiral galaxy (Hidaka,
Sofue 2002).

(a) (Top left:) A wind with a gas density of $5\times 10^{-4}$ H cm$^{-3}$  
blows from the west (right) to east (left) at an inflowing angle of 
45$^\circ$ with a wind velocity of 1000 \kms.
The rotation of the galaxy is anticlockwise, just like NGC 4254.
Note that the faint outer features are artifact due to an edge-reflection
effect in the calculation.
(b) (Top right:) The same, but with a wind velocity of 1500 \kms.
(c) (Bottom:) Observed CO intensity distribution in NGC 4254
(same as figure 4a).

\end{document}